\documentclass[11pt,a4paper]{article}
\pdfoutput=1
\usepackage{jcappub}
\usepackage[english]{babel}
\usepackage{xspace} 
\usepackage[tight]{subfigure} 
\usepackage{url}

\usepackage{cosmodefs.JCAP}

\usepackage{amsmath}
\usepackage{amssymb}
\usepackage{ifthen,xcolor}
\begin{document}

\date{\today}
\title{A Smooth Landscape: Ending Saddle Point Inflation Requires Features to be Shallow}
\author{Diana Battefeld$^{1)}$}
\emailAdd{dbattefe(AT)astro.physik.uni-goettingen.de}
\author{Thorsten Battefeld$^{1)}$}
\emailAdd{tbattefe(AT)astro.physik.uni-goettingen.de}
\affiliation{1) Institute for Astrophysics,
University of Goettingen,
Friedrich Hund Platz 1,
D-37077 Goettingen, Germany}

\abstract{We consider inflation driven near a saddle point in a higher dimensional field space, which is the most likely type of slow roll inflation on the string theoretical landscape; anthropic arguments need to be invoked in order to find a sufficiently flat region. To give all inflatons large masses after inflation and yield a small but positive cosmological constant, the trajectory in field space needs to terminate in a hole on the inflationary plateau, introducing a curved end-of-inflation hypersurface. We compute non-Gaussianities (bi- and tri-spectrum) caused by this curved hyper-surface and find a negative, potentially large, local non-linearity parameter. To be consistent with current observational bounds, the hole needs to be shallow, i.e.~considerably wider than deep in natural units. To avoid singling out our vacuum as special (i.e.~more special than a positive cosmological constant entails), we deduce that all features on field space should be similarly shallow, severely limiting the type of landscapes one may use for inflationary model building. 

We  justify the use of a truncated Fourier series with random coefficients, which are suppressed the higher the frequency, to model such a smooth landscape by a random potential, as is often done in the literature without a good a priory reason.  }

\maketitle

\section{Introduction}
The presence of a plethora of metastable vacua on the string theory landscape \cite{Bousso:2000xa,Susskind:2003kw,Douglas:2006es}\footnote{See \cite{Copsey:2013jla} for arguments questioning the stability of orientifolds and thus the existence of the landscape.}, has been used to explain the small but positive value of the cosmological constant by means of anthropic arguments \cite{Weinberg:1987dv,Bousso:2008bu}. Such arguments gained traction, since the same landscape also provides a dynamical mechanism to scan these vacua dynamically  during eternal inflation \cite{Brown:2011ry}. However, quantitative studies remain speculative due to the measure problem (see e.g.~\cite{Olum:2012bn,Schiffrin:2012zf} for a recent analysis of the problem and \cite{Freivogel:2011eg} for a review of proposed measures), as well as our poor understanding of vacua lifetimes on higher dimensional field spaces: several studies indicate that the lifetime of metastable vacua decreases dramatically through various effects as the dimensionality of field space is increased \cite{Tye:2007ja,Sarangi:2007jb,Podolsky:2007vg,Podolsky:2008du,Brown:2010bc,Brown:2010mg,Brown:2010mf,Greene:2013ida}; in addition, getting trapped classically in a metastable vacuum with positive energy becomes increasingly unlikely  \cite{Battefeld:2012qx}.

In this study we take as a working hypothesis that the basic idea of a landscape leading to eternal inflation (see e.g.~\cite{D'Amico:2012ji} for a recent proposal) and an anthropic explanation of the cosmological constant is correct. Since tunnelling out of an eternally inflating metastable vacuum via a Coleman-deLucia instanton \cite{Coleman:1980aw} leads to an open universe in the nucleated bubble \cite{Bucher:1994gb,Linde:1995xm,Yamamoto:1995sw,Linde:1995rv,GarciaBellido:1997te} (see e.g.~\cite{Sugimura:2011tk,Battefeld:2013xka} for a concrete two-field setup), around sixty e-folds of slow roll inflation are needed subsequently to dilute curvature to levels that do not impede the formation of gravitationally bound objects such as galaxies \cite{Vilenkin:1996ar,Garriga:1998px,Freivogel:2005vv} (see \cite{Guth:2012ww} for using the observation of curvature or lack thereof as a test of the landscape idea); such reasoning invokes anthropics. In a realistic landscape, many contributions, set by the concrete compactification of the Calabi-Yau space, determine the effective, four-dimensional potential. As a consequence, the potential may often be approximated by a random potential, subject to certain constraints set by the type of moduli space under consideration, e.g.~the Denev-Douglas landscape \cite{Denef:2004cf} has been modelled by a random potential \cite{Marsh:2011aa}(see also \cite{Chen:2011ac,Bachlechner:2012at}). 
A further example is the KKLMMT \cite{Kachru:2003sx}
brane inflation proposal \cite{dvali-tye,Alexander:2001ks,Burgess:2001fx,Dvali:2001fw, Firouzjahi:2003zy,Burgess:2004kv,Buchel,Iizuka:2004ct} with turned on angular directions: the moduli acquire a complicated potential that can be approximated by a specific class of random potentials \cite{Agarwal:2011wm}, which can be used to make predictions for inflation in this particular landscape \cite{McAllister:2012am,Dias:2012nf}\footnote{Discrepancies of the results in \cite{Dias:2012nf} with \cite{McAllister:2012am} are explained in \cite{McAllister:2012am}.}); predictions can be made either numerically, by investigating a large number of potentials, or analytically based on random matrix theory. See also \cite{Linde:2007jn,BlancoPillado:2012cb,Sumitomo:2012cf,Sumitomo:2012vx} for related work.

In such random potentials  small-field inflation near a saddle point is far more likely than large-field inflation \cite{Frazer:2011tg,Frazer:2011br,Agarwal:2011wm,Battefeld:2012qx,Pedro:2013nda}; further, inflation driven near a saddle is of short duration (the probability scales usually as $N^{-3}$ \cite{Freivogel:2005vv,Agarwal:2011wm,Yang:2012jf}, where $N$ is the number of e-folds) and driving the last sixty e-folds of inflation at a single saddle is considerably more likely \cite{Aazami:2005jf} than stringing together several short bursts of inflation at distinct saddles, as envisioned in folded inflation \cite{Easther:2004ir}. The encounter with an appropriately flat plateau is unlikely, so that anthropic arguments need to be invoked. 

If inflation is terminated  by falling off a saddle, the most likely final resting place would provide a large negative contribution to the cosmological constant if the potential is not sharply bounded from below at $V=0$ and the dimensionality of field space is large \cite{Battefeld:2012qx}; by large we mean that three or more fields are light on the inflationary saddle\footnote{In the landscape of \cite{Pedro:2013nda}, $n_{\text{total}}\geq 8$ is needed for $n_{\text{inf}}\geq 3$ light fields to become likely.}. To avoid this problem, one needs to invoke anthropic arguments to argue that the trajectory falls into a hole right next to the inflationary plateau with a small, but positive vacuum energy at its bottom and steep walls such that all inflaton fields are stabilized  subsequently \cite{Battefeld:2012qx} (similar to a hole on a golf course). Due to the strong reliance on anthropic arguments to single out the inflationary history during the last sixty e-folds of inflation we refer to this setup as anthropic saddle point inflation.

In this article, we compute the generation of non-Gaussianities caused by the end of inflation, i.e. the generation of a bi- and tri-spectrum caused by the curved end-of-inflation hyper-surface given by the boundary of the hole.

We expand the potential near the saddle point and assume a straight trajectory during inflation to isolate effects originating at the end of inflation  \cite{Huang:2009vk} from contributions to non-Gaussianities stemming from a curved  inflationary trajectory -- the latter ones are well understood and usually lead only to unobservable small non-Gaussianities \cite{Battefeld:2006sz} \footnote{Non-Gaussianities generated during saddle-point inflation are often not frozen, since the adiabatic regime has not been reached yet \cite{Battefeld:2009ym,Elliston:2011dr}; adiabaticity can be achieved by entering a narrow valley during inflation, at which point an EFT \cite{Achucarro:2010jv,Achucarro:2010da,Cespedes:2012hu,Achucarro:2012sm,Achucarro:2012yr,Achucarro:2012fd} 
may be used \cite{Cespedes:2013rda}, or during reheating \cite{Leung:2013rza}; in either case, non-linearity parameters need to be evolved until such a regime is entered, before a comparison with observations can be made. However, the prior amplitude can be used as an order of magnitude estimator \cite{Leung:2013rza}.} particularily for inflation on random potentials \cite{Frazer:2011br,McAllister:2012am}. We assume that the degree of freedom along the trajectory dominates contributions to the power-spectrum, such that parameters in the potential can be computed by comparison with observations, Sec.~\ref{sec:singlefieldinflation}. We justify this assumption subsequently by considering observational constraints on non-Gaussianities \cite{Ade:2013tta}. We find that all but one parameter, the inflationary energy scale $C_0$, are fixed in this manner, whereas the latter is bounded from above by the requirement of small-field inflation. We provide simple analytic approximations expressing all model parameters in terms of $C_0$ and observables. This prototypical model of small field inflation is well known, see e.g.~\cite{Hotchkiss:2011am}, but the particularly simple analytic relationships have not been stated explicitly elsewhere to our knowledge.

We use this simple inflationary model to compute analytically the local non-linearity parameter $f_{NL}$ of the bi-spectrum caused by the curved end-of-inflation hypersurface \cite{Huang:2009vk} in Sec.~\ref{sec:bi-spectrum}, using the $\delta N$-formalism \cite{Starobinsky:1986fxa,Sasaki:1995aw,Rigopoulos:2003ak,Lyth:2004gb,Lyth:2005fi}. The parameters of the tri-spectrum, $\tau_{NL}$ and $g_{NL}$, are provided in App.~\ref{app:trispectrum}.  We retain only the leading order contribution originating from the field in whose direction the curvature of the hyper-surface is largest. 

We interpret our results in Sec.~\ref{sec:discussion}: due to the convex end-of-inflation hypersurface, $f_{NL}$ is always negative if it is of observable size. The magnitude of non-Gaussianities depends on the impact parameter $\mu$, parametrizing the angle at which the hole is hit, the radius of the hole $R$ and in certain cases the inflationary energy scale. This magnitude  grows as the radius of the hole decreases. If the width of the hole is comparable to it's depth in natural units, $f_{NL}$ is large and negative for the vast majority of impact parameters, even for the most favourable inflationary energy scales. Since anthropic arguments do not favour small non-Gaussianities (as indicated by current experiments such as PLANCK \cite{Ade:2013tta}), they can not be invoked to argue for a finely tuned impact parameter $\mu/R\ll 1$. We conclude that the encounter with such a hole violates current observations and is thus ruled out. A shallower hole, e.g.~$R\sim \sqrt{C_0}$, reduces $|f_{NL}|$ sufficiently for the majority of impact parameters, rendering such an end of inflation viable.

Based on the Copernican principle, we argue that the properties of the hole should be representative of the landscape, indicating a sufficiently smooth potential such that the extent of features is considerably larger than their height in natural units. One might say that the potential on the field space resembles a beginners ski slope, without sharp cliffs or mountains. This in turn justifies  the use of a truncated series expansion (with suppressed coefficients the higher the frequency) to model the landscape by a smooth random potential,  as in  \cite{Tegmark:2004qd,Frazer:2011tg,Frazer:2011br,Battefeld:2012qx}.

\section{Inflation near a Saddle Point}
On random potentials, inflation occurs most often at or near a single saddle point \cite{Frazer:2011tg,Frazer:2011br,Battefeld:2012qx,Agarwal:2011wm}. While the trajectory is generically curved, this bending only leads in rare cases to observable non-Gaussianities \cite{Frazer:2011br,McAllister:2012am}; here, we are primarily interested in the effects of the end of inflation onto observables. To isolate these effects, we approximate the inflationary trajectory  by a straight line in field space traversing an inflection point, with at least one additional light perpendicular field.

At this point the reader may wonder why additional light fields should be present during saddle point inflation. Their presence or absence is contingent on the type of landscape under consideration. However, in a higher dimensional random potential (we are thinking of the order of $D\sim \mathcal{O}(100)$ fields or more as motivated by the string theoretical landscape) the presence of several light fields near a saddle point is well motivated based on results of random matrix theory: the joint probability distribution of eigenvalues of the Hessian at a critical point can usually be approximated by Wigner's semicircle law due to the feature of universality in random matrix theory (see \cite{Battefeld:2012qx} for a brief overview of relevant results as well as extensive references to the original literature); one can visualize the distribution of eigenvalues of the Wigner ensemble by means of a Coulomb gas \cite{Dyson:1962.1,Dyson:1962.2}: the eigenvalues correspond to the positions of $D$ charged beads undergoing Brownian motion on an infinite string, with a confining quadratic potential at the origin and logarithmic repulsion between the beads. The probability of finding positive/negative eigenvalues corresponds to the probability of finding beads left/right of the origin. It is clear that saddles with only one unstable direction (negative eigenvalue), i.e. an inflationary valley, are rarer than saddles with several unstable directions. Of course, as more unstable directions enter, it becomes more likely that at least one of them has a large negative square mass. As a consequence, inflation on that saddle would not last long. Thus, there is a preference for having a few light fields on an inflationary saddle if $D$ is large and inflation should last for at least sixty e-folds. In the following, we assume that at least one perpendicular light field is present.

\begin{figure*}[tb]
\begin{center}
\includegraphics[scale=0.5,angle=0]{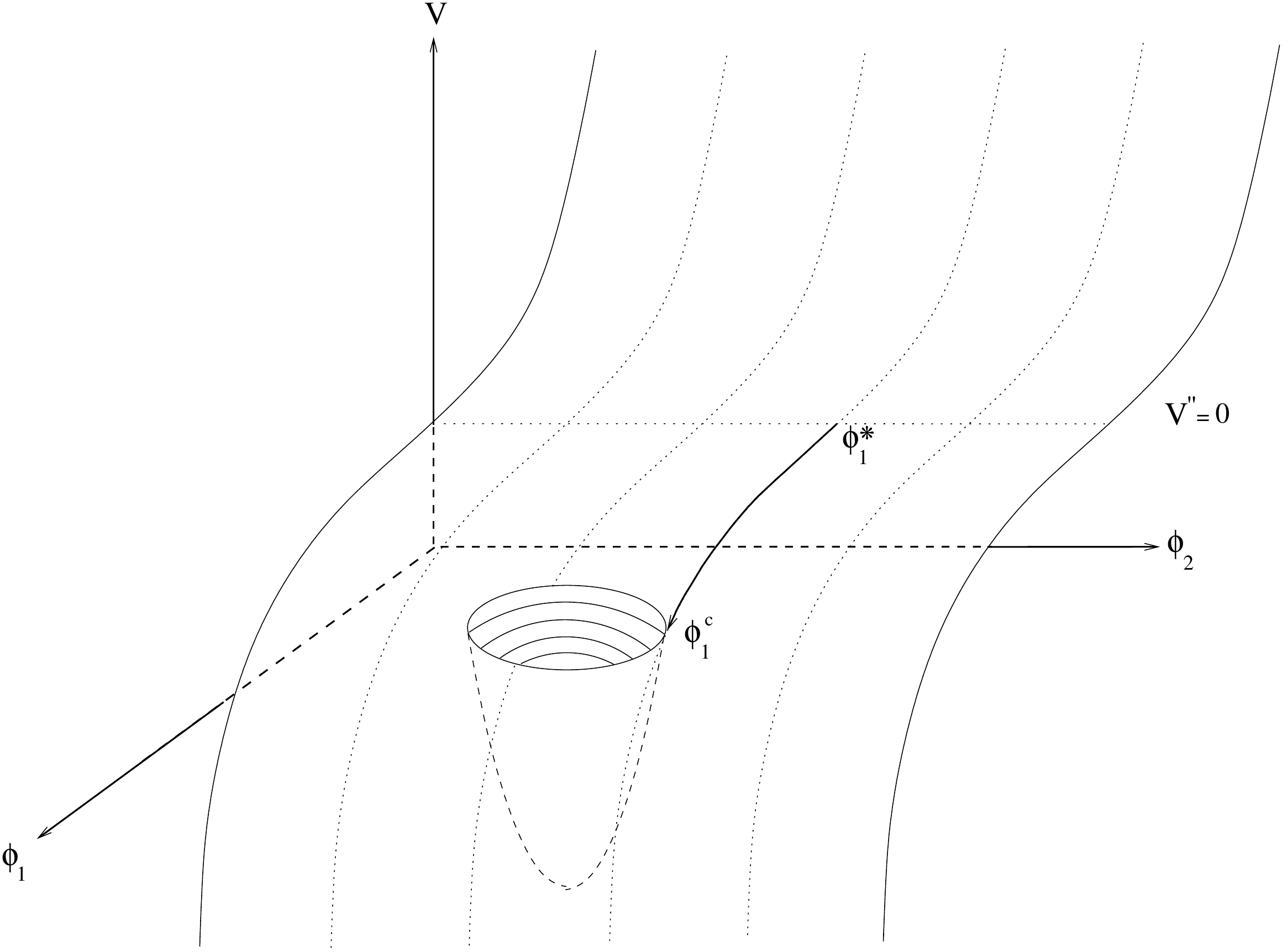}
   \caption{Schematic of small-field inflation near a saddle: $\phi_1$ is the inflaton, while $\phi_2$ is representative of perpendicular light fields. Inflation takes place on the plateau; $\phi_1^*$ denotes the field value at $N=60$ e-folds before the end of inflation, when observable scales left the horizon (a red spectral index requires $\phi_1^*$ to lie on the lower side of the saddle where $|V^\prime|=|\partial V/\partial \phi_1|$ is increasing in the direction of the trajectory). Inflation is terminated when the trajectory falls into the hole, whose minimum corresponds to the current value of the cosmological constant. The expansion of the potential in (\ref{potential}) is valid on the plateau, whereas the potential in the hole is much steeper. See Fig.~\ref{pic:trajectory2} for more details.  \label{pic:trajectory} }
   \end{center}
\end{figure*}

Small-field inflation may thus occurs on this plateau, but the potential needs to change abruptly to terminate inflation: in order to end up in a vacuum  with large masses for all inflatons, the trajectory must lead into hole, similar to the hole on a golf course, see Fig.~\ref{pic:trajectory}. To yield a positive value of the cosmological constant, the  height of the hole has to be finely tuned, particularly if the random potential allows for negative vacua, i.e.~it is not sharply bounded from below at zero  -- in \cite{Battefeld:2012qx}, we showed that the probability of finding a positive, metastable vacuum directly after inflation near a saddle point on a random potential is strongly suppressed, the more so the higher the dimensionality of field space. To reconcile the observation of a positive cosmological constant with the low probability of finding such a vacuum by chance, the anthropic principle needs to be invoked; alternatively, one may conclude the the proposal of a such random landscapes with several light fields during inflation is ruled out. 

Whether or not the hole lies on the inflationary plateau or is reached after a subsequent phase of fast roll expansion, which may involve a strongly curved trajectory, depends on the type of landscape one considers. In the smooth landscapes we considered in \cite{Battefeld:2012qx} (given by a truncated Fourier series), trajectories after inflation near a saddle point where indeed curved, but they also lead to large negative values of the cosmological constant. Since the field can not traverse a long distance in field space during fast roll without leading to large negative values of the potential, it appears plausible to us that invoking a prior on positive values of the cosmological constant singles out trajectories that fall preferably into a hole on the inflationary plateau. This conjecture needs to be tested in concrete landscapes, which we plan to address in a future publication; for now we wish to focus on a hole located on the inflationary plateau\footnote{It should be noted that some predictions, such the amplitude and sign of the bi-spectrum, can depend on post-inflationary dynamics, such as the presence of a winding trajectory in a gorge before a hole is reached. We comment on this possibility in Sec.~\ref{sec:discussion}.}. 

Thus, in this article we take as working hyposthesis that the landscape idea combined with anthropic reasoning is correct, leaving aside the issues raised by the measure problem \cite{Olum:2012bn,Schiffrin:2012zf,Freivogel:2011eg} or the potentially short lifetime of metastable vacua on higher dimensional field spaces \cite{Tye:2007ja,Sarangi:2007jb,Podolsky:2007vg,Podolsky:2008du,Brown:2010bc,Brown:2010mg,Brown:2010mf,Greene:2013ida}. Our goal is to investigate if this peculiar type of anthropic saddle point inflation is consistent with observations.

To this end, we use a simple small field model near a saddle point for inflation, consistent with observational constraints \cite{Ade:2013lta,Hinshaw:2012fq,Bennett:2012fp,Desjacques:2009jb,Vielva:2009jz,Smidt:2010sv,Smidt:2010ra}, and examine the additional signatures caused by the curved end-of-inflation hyper-surface in field space; this  effect is generic for multi field inflation as long as the inflationary trajectory is not perpendicular to the end-of-inflation hypersurface  \cite{Huang:2009vk}. We are particularly interested in non-Gaussianities, which are sensitive to small changes in the volume expansion rate caused by isocurvature fluctuations towards the end of inflation. Throughout this article, we set
\begin{eqnarray}
M_P^2=\frac{1}{8\pi G}\equiv 1\,.
\end{eqnarray}

\subsection{The Inflationary Phase \label{sec:singlefieldinflation}}
 
\begin{figure*}[tb]
\begin{center}
\includegraphics[scale=0.5,angle=0]{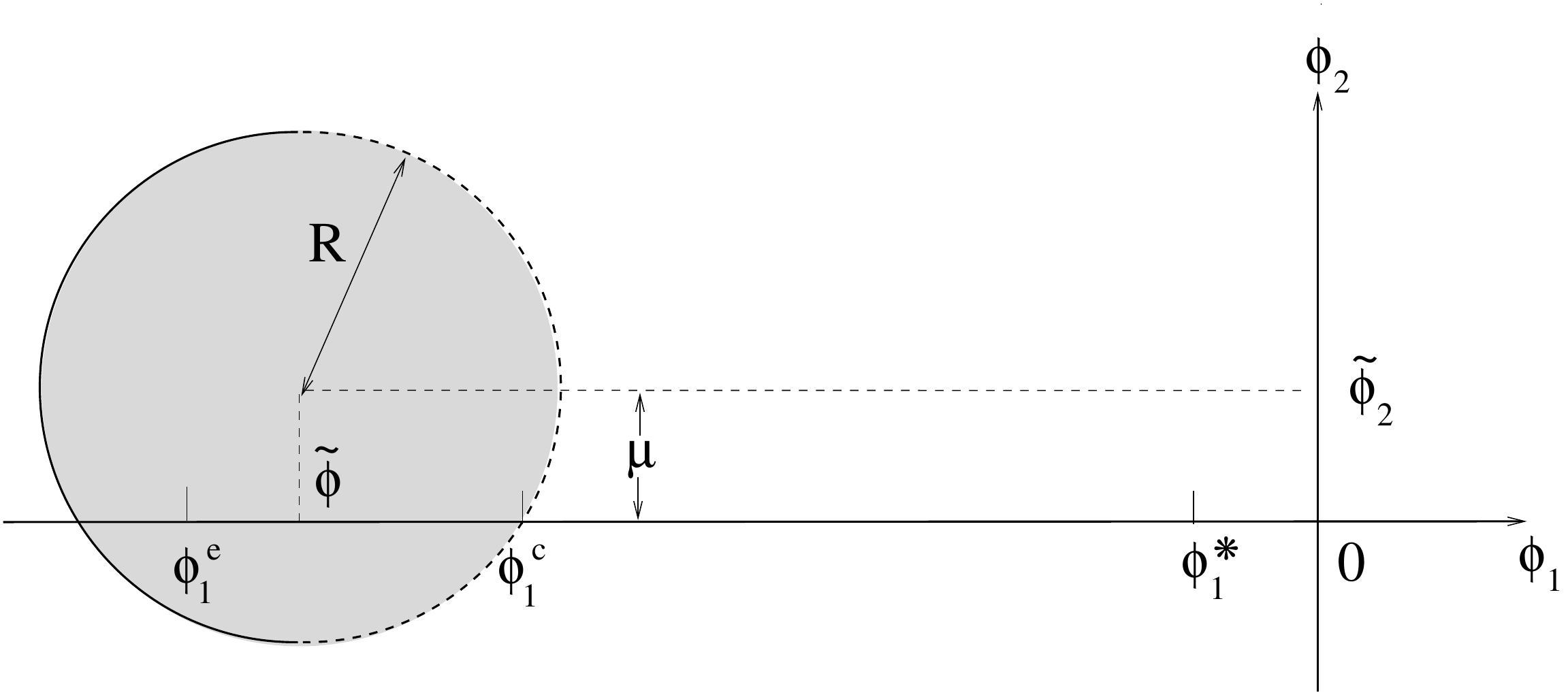}
   \caption{Top view of the hole (shaded circle) with radius $R$ on the inflationary plateau and origin at $\phi_i=\tilde{\phi}_i$ (i.e.~the minimum of $V(\varphi)$): the $\phi_1$--$\phi_2$ plane contains the tangent to the inflationary trajectory when the hole is encountered at $\vec{\phi}^c$ as well as the origin of the hole;
the dashed semicircle, representing the right boundary of the hole, is part of the intersection of the curved end-of-inflation hypersurface with the $\phi_1$--$\phi_2$ plane, which we denote by $\phi_1(\phi_2)\equiv g(\phi_{2})$. The curvature of $g$ is largest in this plane; as a consequence, we focus on the additional effects brought forth by $\phi_2$ in this paper. Coordinates are chosen such that $\phi_2=0$ during inflation and $\phi_1=0$ at $V^{\prime\prime}=0$. $\phi_1^e<\phi_1^c$ indicates the field value at which slow roll would break down in the inflationary potential (\ref{potential}) if the hole were absent. $\mu$ denotes the impact parameter.
  \label{pic:trajectory2}}
   \end{center}
\end{figure*}

Consider a multi-dimensional field space  with a potential that reduces to
\begin{eqnarray}
V(\phi_1,\dots,\phi_n )|_{\text{infl.}}=V(\phi_1)
\end{eqnarray}
during inflation and
\begin{eqnarray}
V(\phi_1,\dots, \phi_n)|_{\text{reh.}}&=&V(\varphi)\,,
\end{eqnarray}
with
\begin{eqnarray}
\varphi&\equiv& \sqrt{\sum_{i=1}^{n}(\phi_i-\tilde{\phi}_i)^2}\quad , \quad \tilde{\phi}_i=\text{const.}
\end{eqnarray}
and $V(\varphi=0)=V_{\text{c.c.}}$, once the trajectory traverses the (sharp) boundary to a hole. We keep the hypersurface designating the end of inflation, 
\begin{eqnarray}
\phi_1(\phi_2,\dots,\phi_n)\equiv g(\phi_2,\dots, \phi_n)\,,
\end{eqnarray}
general for the most part of this study, but have a hyper-sphere in mind: a hole with radius $R$ and minimum at $\vec{\phi}_h=(\tilde{\phi}_1,\dots,\tilde{\phi}_n)$. 
We denote field values at the end of inflation\footnote{In our notation a '$_*$' denotes an initial flat hypersurface and a '$_c$'  a (final) uniform energy density hypersurface to enable the use of the non-linear $\delta N$ formalism.} by $\vec{\phi}_{c}\equiv (\phi_1^c,\dots,\phi_2^c)=(g(\phi_2^c,\dots,\phi_n^c),\phi_2^c,\dots,\phi_n^c)$.
Without loss of generality, we define $\phi_2$ to be the field perpendicular to $\phi_1$, such that the plane spanned by $\phi_1$ and $\phi_2$ contains the tangent to the inflationary trajectory at $\vec{\phi}_c$ as well as the minimum of the hole at $\vec{\phi}_h$ \footnote{We are not interested in the fine tuned limit $\mu\rightarrow 0$ where this definition becomes ill defined (note that non-Gaussianities vanish in this limit).}. Thus, we may define an impact parameter as
\begin{eqnarray}
\mu\equiv \tilde{\phi}_{2}\leq R\,,
\end{eqnarray}
where $R$ designates the radius of the hole, see Fig.~\ref{pic:trajectory2}.

We further assume that inflation is driven near an inflection point where $\partial^2 V/\partial\phi_1^2\equiv V''=0$ and  $V'$ is small; without loss of generality, we can set $\phi_1=0$ when $V''=0$ and expand the inflationary potential  as
\begin{equation}
V(\phi_1)=C_0+C_1\phi_1+C_3\phi_1^3\,,\label{potential}
\end{equation}
where $C_0$, $C_1$ and $C_3$ are constants to be determined by comparison with observations of e.g.~the cosmic microwave background radiation \cite{Hinshaw:2012fq,Bennett:2012fp,Ade:2013lta}. We require
\begin{itemize}
\item at least sixty e-folds of inflation, $N=60$,
\item small field inflation, $\Delta \varphi_1 < 1$, consistent with suppressed gravitational waves (the current bound \cite{Ade:2013lta} on the tensor to scalar ratio is $r_{0.002}<0.11$),
\item an amplitude of the power-spectrum satisfying the COBE-bound \cite{Hinshaw:2012fq,Bennett:2012fp}, $P_\zeta^{\text{COBE}}=(2.43\pm 0.11)\times 10^{-9}$,
\item a red scalar spectral index \cite{Ade:2013lta} 
$n_s=0.9603 \pm 0.0073$ 
\end{itemize}
These constraints fix two out of the three parameters in the potential, while imposing a strict upper bound on the third.
Since we assume a flat potential in the perpendicular directions, we have
\begin{eqnarray}
\phi_i^*=\phi_i^c\equiv 0\quad \text{for}\quad i=2,\dots,n \,,
 \end{eqnarray} 
and can therefore focus entirely on $\phi_1$ during inflation (see e.g.~\cite{Bassett:2005xm} for a review of inflationary slow roll dynamics). 
We further assume that the dominant contribution to the power-spectrum stems from fluctuations in $\phi_1$, i.e.~isocurvature fluctuations in the perpendicular fields transferred at end of inflation are allowed to dominate higher order correlation functions, but not $P_\zeta$ (we shall see in Sec.~\ref{sec:discussion} that the dominant contribution can not originate from $\phi_2$ height-to-width ratio ($R\propto \sqrt{C_0}$) due to current bounds on the bi-spectrum).

Since we are primarily interested in the order of magnitude of parameters, we work in the slow roll limit: we recover the results of \cite{Hotchkiss:2011am}, providing additional analytic insights.
The slow roll parameters $\epsilon\equiv(V'/V)^2/2$ and $\eta\equiv V''/V$ are
\begin{eqnarray}
\epsilon(\phi_1)&=&\frac{1}{2}\left[\frac{C_1^2+6C_1C_3\phi_1^2+9C_3^2\phi_1^4}{C_0^2+2C_0C_1\phi_1+C_1^2\phi_1^2+2C_0C_3\phi_1^3+2C_1C_3\phi_1^4+C_3^2\phi^6_1}\right]\,,\\
\eta(\phi_1)&=&\frac{6C_1C_3\phi_1}{C_0+C_1\phi_1+C_3\phi_1^3}\,.
\end{eqnarray}
In hindsight, observational constraints entail that
 $C_0\gg C_1|\phi_1^*|,C_3|\phi_1^{*3}|$, $C_3/C_1\gg (n_s-1))\sim10^{-2}$, $C_1^2/C_0^2\ll (n_s-1)\sim10^{-2}$ and $C_1^3/(C_0^2C_3)\ll1$, so that we can approximate the slow roll parameters when observable modes crossed the horizon at $\phi_1^*$ by
\begin{eqnarray}
\epsilon(\phi_1^*)&\approx &\frac{1}{2}\left[\frac{C^2_1+6C_1C_3\phi_1^2}{C_0^2}\right]_*\,,\label{epsilonapprox}\\
\eta(\phi_1^*)&\approx & \frac{6C_3\phi_1}{C_0}\bigg|_*\,,\label{eta}
\end{eqnarray}
where we only kept the leading order contributions. Note that these expressions are not simple Taylor expansions around $\phi_1=0$, since a red spectral index requires a small but non-zero $\phi_1^*\neq 0$ and coefficients in the potential are very small. The above approximations are self consistent.

Let us first compute the initial value, $\phi_1^*$. Within the slow roll approximation, the scalar spectral index becomes
\begin{equation}
n_s\simeq 1+2\eta^*-6\epsilon^*\,.\label{ns}
\end{equation}
Since we deal with small field inflation, $\epsilon \ll \eta$, we can approximate 
\begin{equation}
n_s\approx 1+2\eta^*\,.
\end{equation}
Using (\ref{eta}) we get
\begin{equation}
\phi^*_1\approx \frac{C_0}{6C_3}\frac{n_s-1}{2}\,. \label{phi1*}
\end{equation}
Note that $\phi_1^*<0$ to yield a red spectral index. As a consequence, the last sixty e-folds of inflation are preceded by a similar amount of slow roll inflation with $\phi_1>0$. 

At this point we would like to comment on the role of tuning: since there is no anthropic reason to prefer a red tilt over a blue one, we need to rely on fine tuning to reconcile inflation at a saddle point with the observation of a red tilt by PLANCK at the 5$\sigma$ level. Since long burst of inflation are less likely than short ones, e.g.~$P\propto N^{-3}$ in \cite{Freivogel:2005vv,Agarwal:2011wm,Yang:2012jf}, one needs to invoke tuning at the percent level; hence, this type of inflation is already disfavoured\footnote{This statement depends on the choice of prior \cite{Freivogel:2005vv} as well as measure (whether or not additional inflation should be rewarded; here, we do not reward extra inflation as in the uniform scale factor measure).}. However, given that small field inflation is apparently well motivated  by the analysis of the PLANCK data as well as  conceptually by the landscape idea, we will be content with tuning at the percent level in this study as long as no additional fine tuning is needed.

If inflation were to be terminated via the breakdown of slow roll, $|\eta|\sim 1$, in the potential (\ref{potential}), it would end at 
\begin{equation}
\phi_1^{\text{e}}=-\frac{C_0}{6C_3}\,,
\label{inflaton}
\end{equation}
requiring appropriate adjustments of parameters in the potential to yield sixty e-folds of inflation. However, we are interested in placing a hole at $\phi_1^c>\phi_1^e$ on the plateau. Thus, we need to make sure that the plateau supports more than sixty e-folds of inflation for $\phi_1<0$ if the hole were to be missed. Therefore we demand
\begin{eqnarray}
100\equiv N=\int_*^e H\, dt
\simeq -\int_*^e\frac{V}{V'}d\phi_1=-\frac{{C_0}}{\sqrt{3C_1C_3}}\arctan\left[\phi_1\sqrt{\frac{3C_3}{C_1}}\right]_*^e\,.\label{efolds}
\end{eqnarray}
Approximating $\phi_1^*\approx 0$ in this expression (a poor approximation \footnote{Using the full volume expansion rate, as we will do in Sec.~\ref{sec:curvedhypersurface}, with the values of $C_i$ derived in this section leads to $N\approx 63$ instead of the prescribed $N\equiv 100$; we chose the prescribed value above, since we want to put the encounter with the hole at $60$ e-folds after $t=t_*$  and did not want to demand a much flatter plateau than actually needed.} in our case, leading to an error of about $40$\% in $N$, but enabling a simple analytic understanding of model parameters)  as well as $\arctan(x)\approx -\pi/2-1/x$ (this is valid for large, negative $x$), we get
\begin{eqnarray}
N\approx \frac{C_0\pi}{2\sqrt{3C_1C_3}}-2\,,\label{reducedN}
\end{eqnarray}
leading to
\begin{equation}
C_1\approx \frac{C_0^2\pi^2}{12(N+2)^2C_3}\,. \label{C1}
\end{equation}

To express $C_3$ in terms of $C_0$ and observables, we use the amplitude of the power-spectrum in the slow roll approximation, 
\begin{equation}
P_\zeta\simeq \frac{42}{8^3\pi^2}\frac{C_0}{\epsilon^*} \,,\label{PR}
\end{equation}
and solve for 
\begin{eqnarray}
\epsilon^*\simeq \frac{42}{8^3\pi^2}\frac{C_0}{P_\zeta}\,,\label{epsilonobs}
\end{eqnarray} 
which can be used in (\ref{epsilonapprox}) with (\ref{phi1*}) to yield
\begin{equation}
C_3\approx \frac{C_0}{12\sqrt{2\epsilon^*}}\left[\frac{\pi^2}{(N+2)^2}+\left(\frac{n_s-1}{2}\right)^2\right]\,.\label{C3}
\end{equation}
Thus, the only free parameter left is $C_0$. It's value is bounded from above by the requirement of small-field inflation,
\begin{eqnarray}
1=\Delta\phi_\text{max}\geq \Delta \phi_1=|\phi_1^c-\phi_1^*|\sim |\phi_1^e|=\frac{C_0}{6C_3}\,.
\end{eqnarray}
Using (\ref{C3}) with (\ref{epsilonobs}) in this expression leads to
\begin{eqnarray}
C_0&\leq& \frac{8^3\pi^2P_\zeta}{84}\left(\frac{\Delta \phi_{\text{max}}}{2}\right)^2\left[\frac{\pi^2}{(N+2)^2}+\left(\frac{n_s-1}{2}\right)^2\right]^2\\
&\approx& 0.66\times 10^{-13}
\equiv C_0^{\text{max}}\,\label{C0max}.
\end{eqnarray}
Taking values of $C_0$ below this limit corresponds to decreasing $\Delta \phi_1$ below $\Delta \phi_{\text{max}}$, and thus additional (unnecessary) fine tuning (see also \cite{Pedro:2013nda}, where it is argued that higher inflationary scales are more likely). Changing $C_0$ corresponds to moving along the green central contour in Fig.~1 of \cite{Hotchkiss:2011am}. 
Using (\ref{C1}) and (\ref{C3}), the corresponding limits for $C_1$ and $C_3$ are
\begin{eqnarray}
C_1&\leq& 0.31\times 10^{-16}\equiv C_1^{\text{max}}\,,\\
C_3&\leq& 1.1\times 10^{-14}\equiv C_3^{\text{max}}\,.
\end{eqnarray}
At this stage, one may easily check that the approximations we have made to simplify the slow roll parameters in (\ref{epsilonapprox}) and (\ref{eta}) are valid.  We would like to emphasize that the above model of inflation is well known \cite{Hotchkiss:2011am} and serves merely as a consistent, simple backdrop for us to investigate the effects caused by the peculiar end of inflation. The error of about $40\%$ caused by ignoring $\phi_1^*$ in the volume expansion rate above is irrelevant for our subsequent conclusions. 

\subsection{Ending Inflation on a Curved Hypersurface \label{sec:curvedhypersurface}}

During inflation, perturbations in perpendicular, light fields $\phi_i$, $i\geq 2$, acquire a nearly scale invariant spectrum of fluctuations, like the inflaton. If the surface at which inflation ends is curved, the associated isocurvature perturbations yield a local variation of the volume expansion rate $\delta N$, which in turn feeds into the curvature fluctuation $\zeta$; the latter can be computed via the $\delta N$-formalism. This basic mechanism is relatively simple and known in the literature \cite{Huang:2009vk}, but its application to the present setup is novel.

In our case, we are motivated by inflation in random potentials \cite{Tegmark:2004qd,Frazer:2011tg,Frazer:2011br,Battefeld:2012qx}, where the encounter of a flat region is rare, requiring the use of anthropic arguments. Further anthropic arguments are needed to terminate inflation in a vacuum that stabilizes all fields while yielding a small positive cosmological constant \cite{Battefeld:2012qx}; in its simplest realization, a hole in which the trajectory ends is located directly next to the plateau, see Fig.~\ref{pic:trajectory}. If the dimensionality of field space is high and values of the potential are not bounded sharply by zero, encounters with such vacua are exceedingly rare even compared with the already rare encounter with a saddle point. Since eternal inflation is an integral part of inflation on the landscape, many trajectories are sampled in the multiverse and therefore anthropic reasoning can be invoked: to dilute curvature and allow for the formation of gravitationally bound structures, subsequent slow roll inflation needs to last about sixty e-folds \cite{Vilenkin:1996ar,Garriga:1998px,Freivogel:2005vv},  and the resulting cosmological constant needs to be small \cite{Weinberg:1987dv,Bousso:2008bu}. Of course, the measure problem \cite{Olum:2012bn,Schiffrin:2012zf,Freivogel:2011eg} arises whenever quantitative, probabilistic predictions are sought after.  We refer the interested reader to our previous article \cite{Battefeld:2012qx} for a more detailed discussion of inflation on random potentials and a summary of anthropic arguments.

If the landscape idea is right, our universe seems to have had a phase of such an anthropically selected saddle point inflation in its past. In this article, we wish to put this framework to the test and ask:

\begin{center}
{\em Are there observational consequences, such as non-Gaussianities, \\caused by the peculiar end of anthropic saddle point inflation?}\\
\end{center}

Since the curvature of $g$ is strongest in the $(\phi_1,\phi_2)$-plane, we focus on the additional (leading order) contributions caused by $\phi_2$, and ignore the effect of $\phi_i$, $i\geq 3$. In this plane, the trajectory encounters the surface $(g(\phi_2^c),\phi_2^c)$ after sixty e-folds of inflation have passed, i.e.~
\begin{eqnarray}
60\equiv N
\simeq -\int_*^c\frac{V}{V'}d\phi_1\approx -\frac{{C_0}}{\sqrt{3C_1C_3}}\arctan\left[\phi_1^c\sqrt{\frac{3C_3}{C_1}}\right]\,,
\end{eqnarray}
yielding
\begin{eqnarray}
\phi_1^c=\sqrt{\frac{C_1}{3C_3}}\tan\left(\arctan\left(\phi_1^*\sqrt{\frac{3C_3}{C_1}}\right)-\frac{N\sqrt{3C_1C_3}}{C_0}\right)
\end{eqnarray}
after specifying $C_0,C_1,C_3$ according to Sec.~\ref{sec:singlefieldinflation}. We do not approximate the volume expansion rate in this section, but take the coefficients in the potential according to (\ref{C1}) and (\ref{C3}). The main effect of using $N=60$ in the full volume expansion rate is a reduction of the previously prescribed $\Delta\phi_{\text{max}}=1$  to
\begin{eqnarray}
\Delta\phi_{\text{max}}\approx -\phi_1^c\approx 0.41\,.
\end{eqnarray} 
given the value of $C_0^{\text{max}}$ in (\ref{C0max}). We use the corrected value of $\Delta\phi_{\text{max}}$, entailing
\begin{eqnarray}
\Delta \phi_1 \approx \Delta \phi_{\text{max}}\sqrt{\frac{C_0}{C_0^{\text{max}}}}\,,
\end{eqnarray}
 in the following. Note that the prescribed values of $n_s-1$ and $P_\zeta$ remain unchanged, as desired.

\subsubsection{The Power-spectrum}
To compute additional contributions to correlation functions, we use the non-linear $\delta N$-formalism \cite{Lyth:2004gb,Lyth:2005fi},
whereby the curvature perturbation on constant energy hypersurfaces on superhorizon scales is related to perturbations in the volume expansion rate
\begin{eqnarray}
\zeta=\delta N=\sum_i \frac{\partial N}{\partial \phi_i}\bigg|_* \delta\phi_i^* +\frac{1}{2}\sum_{i,j} \frac{\partial N}{\partial \phi_i}\frac{\partial N}{\partial \phi_j}\bigg|_*\delta \phi_i^*\delta \phi_j^* +\dots \,.
\end{eqnarray}
The full power-spectrum is
\begin{eqnarray}
P_{\zeta}=\left(\frac{H}{2\pi}\right)^2\sum_{i=1}^{n}\left(\frac{\partial N}{\partial \phi_i}\right)^2\bigg|_{*}\,,
\end{eqnarray}
where
\begin{eqnarray}
H\simeq \sqrt{\frac{V(\phi_1^*)}{3}}\,,
\end{eqnarray}
with $\phi_1^*$ from (\ref{phi1*}). The power-spectrum in (\ref{PR}) is the contribution from $\phi_1$, which we assumed to yield the dominant contribution in Sec.~\ref{sec:singlefieldinflation}.  
 The leading order additional contribution to the power-spectrum  is due to $\phi_2$ and reads
\begin{equation}
\Delta \mathcal P_{\zeta}\simeq \frac{C_0^2}{C_1^2\left(1+g^2\frac{3C_3}{C_1}\right)}\left(\frac{\partial g}{\partial\phi_2^*}\right)^2\frac{H^2}{4\pi^2}\,,
\label{DeltaPzeta}
\end{equation}
where we used\footnote{Here, we use the simplifying assumption that the perpendicular direction is flat; if a small slope is present, one would need to transport the end-of-inflation hypersurface $g(\phi_2^c)$ back in time to $t_*$ yielding $g(\phi_2^*)$; this procedure is straightforward, but requires specifying the potential; further, as long as the slope in the perpendicular direction is shallow and the inflationary trajectory is close to a straight line, which is well motivated for inflation on a saddle, results do not change qualitatively; therefore we will stay with an exactly flat direction in this study.} $N(\phi_1^c)=N(g(\phi_2^c))=N(g(\phi_2^*))$. If the inflationary trajectory is channelled through a valley at some point during the last sixty e-folds of inflation (i.e.~an adiabatic limit is reached before the end of inflation, \cite{Elliston:2011dr}), this contribution would be heavily suppressed (this is not the case for us). To be self consistent, $\Delta P_\zeta$ needs to be sub-dominant, i.e. 
\begin{eqnarray}
\Delta P_\zeta\ll P_\zeta^{\text{COBE}}\,,
\end{eqnarray}
which leads to an upper limit on the slope (in this section a prime denotes a partial derivative w.r.t.~$\phi_2^*$)
\begin{eqnarray}
|g^\prime|
\ll \frac{2\pi}{H}\frac{C_1}{C_0}\sqrt{P_\zeta^{\text{COBE}}\left(1+g^2\frac{3C_3}{C_1}\right)}\,.\label{boundCOBE}
\end{eqnarray}
Using the analytic expressions for the model parameters of Sec.~\ref{sec:singlefieldinflation}, we may write this bound as
\begin{eqnarray}
|g^\prime|&\ll& 13 \,,
\end{eqnarray}
where we used $|g|=|\phi_1^c|\approx \Delta \phi_1= \Delta \phi_{\text{max}} (C_0/C_0^{\text{max}})^{1/2}$, $C_1=C_1^{\text{max}}(C_0/C_0^{\text{max}})^{3/2}$ and $C_3=C_3^{\text{max}}(C_0/C_0^{\text{max}})^{1/2}$. Evidently, the bound on the slope is model independent, i.e.~independent of the inflationary energy scale $C_0$. This simple and useful fact became apparent via the series of approximations made in Sec.~\ref{sec:singlefieldinflation}, and is the main reason we employ them in the first place. For a spherical hole intercepted at $(\phi_1^c,\phi_2^c=0)$ with impact parameter $\mu$ and radius $R$,
\begin{eqnarray}
g(\phi_2)=\sqrt{R^2-(\phi_2-\mu)^2}-\sqrt{R^2-\mu^2}+\phi_1^c\,,\label{defg}
\end{eqnarray}
this bound translates into an upper limit on the impact parameter 
\begin{eqnarray}
\frac{\mu}{R}
\ll \left(1+\frac{1}{|g^\prime|^2}\right)^{-1/2}\equiv  \frac{\mu_{\text{max}}}{R}\approx 0.9972\,,\label{boundonmu}
\end{eqnarray}
ruling out extreme grazing encounters. In addition, we see that tuning of the impact parameter below the percent level would be needed if we wanted to make the contribution to the power-spectrum stemming from $\phi_2$ the dominant one; this would require a readjustment of parameters in the potential to reduce the contribution from $\phi_1$ in order to satisfy the COBE bound. This possibility is well known, and may be used in lieu of the curvaton mechanism. This effect could be included in our study, however, as we will see in the next section, small values of $R$, that is a hole with steep walls, lead to an unacceptable large bi-spectrum for $\mu \sim \mu_{\text{max}}$, so that we do not pursue this avenue.

\subsubsection{The Bi-spectrum \label{sec:bi-spectrum}}
The local limit of higher order correlation functions can be computed via the non-linear $\delta N$-formalism \cite{Lyth:2004gb,Lyth:2005fi}, leading for instance to the local non-linearity parameter of the Bi-spectrum
\begin{eqnarray}
\frac{6}{5}f_{NL}=\frac{N_{i}N_{j}N^{ij}}{(N_k N^k)^2}\,,
\end{eqnarray}
where $N_i$ designates a derivative with respect to $\phi_{i}^*$ etc., and repeated indices are to be summed over the number of fields (see \cite{Suyama:2010uj} for a review). The current observational limit on a local Bi-spectrum is \cite{Ade:2013tta}
\begin{eqnarray}
f_{NL}=2.7\pm 5.8 \label{fnllimits}
\end{eqnarray}
at the $1\sigma$-level, indicating an absence of primordial non-Gaussianities.

Since the contribution of $\phi_1$ is suppressed due to the consistency relation \cite{Maldacena:2002vr}, let us focus on the contribution stemming from $\phi_2$. Taking the second derivative of $N(g(\phi_2^*))$ leads to
\begin{equation}
\frac{6}{5}f_{NL}=\frac{\frac{C_0^3}{C_1^3}\frac{1}{\left(1+g^2\frac{3C_3}{C_1}\right)^3}\left[-\frac{6C_3}{C_1}\frac{1}{\left(1+g^2\frac{3C_3}{C1}\right)}gg^{\prime 2}+g^{\prime\prime}\right]g^{\prime 2}}{\left(P_\zeta\frac{4\pi^2}{H^2}\right)^2}\,,
\end{equation}
where we used that $N_kN^k=P_\zeta(2\pi/H)^2$. Considering the spherical hole defined in (\ref{defg}),
\begin{eqnarray}
g^\prime|_{\phi_{2}^*=0}=\frac{\frac{\mu}{R}}{\sqrt{1-\frac{\mu^2}{R^2}}}\quad , \quad
g^{\prime\prime}|_{\phi_{2}^*=0}=-\frac{1}{R}\left(1-\frac{\mu^2}{R^2}\right)^{-3/2}\,,
\end{eqnarray}
as well as
\begin{eqnarray}
g\approx -\Delta\phi_{\text{max}}\left(\frac{C_0}{C_0^{\text{max}}}\right)^{\frac{1}{2}}\, ,\,
C_1\approx C_1^{\text{max}}\left(\frac{C_0}{C_0^{\text{max}}}\right)^{\frac{3}{2}}\,,\,
C_3\approx C_3^{\text{max}}\left(\frac{C_0}{C_0^{\text{max}}}\right)^{\frac{1}{2}}\,,
\end{eqnarray}
we can simplify the non-linearity parameter to
\begin{eqnarray}
f_{NL}=A_1f_1\left(x\right)-A_2f_2\left(x\right)\left(\frac{C_0}{C_0^{\text{max}}}\right)^{1/2}\frac{1}{R}\label{fNLsimplified}
\end{eqnarray}
where $x\equiv \mu/R$,
\begin{eqnarray}
A_1&\equiv& \frac{5}{(2\pi)^4P_\zeta^2 3^2}
\frac{C_0^{\text{max}\,5} C_3^{\text{max}}\Delta\phi_{\text{max}}}{C_1^{\text{max}\,4}}
\left(1+3\Delta\phi_{\text{max}}^2
\frac{C_3^{\text{max}}}{C_1^{\text{max}}}\right)^{-4}\approx 3.3\times 10^{-10}\,,\\
A_2&\equiv& \frac{5}{6(2\pi)^4P_\zeta^23^2}\frac{C_0^{\text{max}\,5}}{C_1^{\text{max}\,3}}\left(1+3\Delta\phi_{\text{max}}^2
\frac{C_3^{\text{max}}}{C_1^{\text{max}}}\right)^{-3}\approx 6.9\times 10^{-11}\,,
\end{eqnarray}
and
\begin{eqnarray}
f_1(x)\equiv \frac{x^4}{\left(1-x^2\right)^2}\,,\\
f_2(x)\equiv \frac{x^2}{\left(1-x^2\right)^{5/2}}\,.
\end{eqnarray}
This expression is the main result of this paper, whose consequences we explore in the next section.

\section{Discussion: The Smoothness of the Landscape \label{sec:discussion}}
The first term in (\ref{fNLsimplified}) is proportional to $\propto g^{\prime\,4}$ and therefore suppressed due to the constraint in (\ref{boundCOBE}), leading to
\begin{eqnarray}
A_1f_1\left(\frac{\mu}{R}\right)\ll A_1f_1\left(\frac{\mu_{\text{max}}}{R}\right)\approx 1.0\times 10^{-5}\,,
\end{eqnarray}
which is, as expected, unobservable small. 

The second term in (\ref{fNLsimplified}) $\propto g^{\prime\prime}$ is also suppressed, but since it is proportional to $\propto R^{-1}$, it may become large. The function $f_2(x)$ is bounded from above via the bound on $\mu$ in (\ref{boundonmu}) to $f_2(x) \ll f_2(\mu_\text{max}/R)\approx 4.3\times 10^5$, but for impact parameters that are not fine tuned, $x=\mu/R\sim 0.5$, we have $f_2(x)\sim \mathcal{O}(1)$. Further, $C_0$ should be close to $C_0^{\text{max}}$, since a lower value constitutes unnecessary fine tuning from a model builder's perspective.

What are natural values for $R$? The answer to this question depends on the particular landscape under consideration, which we plan to asses in an future publication. Here, we would like to take a phenomenological approach, considering holes with both steep or shallow walls, i.e.~holes with small or large radii respectively.

\subsection{Steep Holes}
Consider a hole that is as wide as it is deep in natural units, i.e.~
\begin{eqnarray}
R\sim C_0\,,
\end{eqnarray}
 such that the second term dominates in $f_{NL}$. 
  The upper bound on $\mu$ translates to $f_{NL}\gg -4.5\times 10^8 $ with $C_0=C_0^{\text{max}}$, while for more natural values of $\mu\sim R/2$, we get 
\begin{eqnarray}
f_{NL}\sim -530\,,
\end{eqnarray}
which is still far outside the allowed interval in (\ref{fnllimits}). Hence, without fine tuning, this particular type of anthropic saddle point inflation is ruled out, casting doubt on the validity of inflationary models on a landscape that incorporate anthropics to yield sufficient inflation near a saddle point and terminate inflation in a hole that is as wide as it is deep.

To reduce the magnitude of $f_{NL}$, the impact parameter could be finely tuned, corresponding to a head on encounter with $\mu/R\ll 1$ (a reduction of $C_0$ does not help, since $f_{NL}\propto 1/\sqrt{C_0}$  due to $R\sim C_0$). Similarly, one may tune the curved hypersurface at which inflation ends such that $g^{\prime\prime}$ is reduced, i.e.~incorporate a more complicated potential towards the end of inflation. However, since there is no anthropic reason why $|f_{NL}|$ should be as small as current observations indicate, and the generic prediction in this class of models ($R\sim C_0\sim V_{\text{inf}}$) is a large negative value, it appears more sensible to us to focus on different scenarios.

\subsection{Shallow Holes}
let us consider a shallower hole, which is still in line with our wish to make all inflatons heavy after inflation;  Taylor expanding around the minimum leads to a quadratic potential   
\begin{eqnarray}
V(\varphi)=\frac{1}{2}m^2\varphi^2\,,
\end{eqnarray}
which we take as the full potential within the hole in this section as an illustrative example.
Large masses correspond to steep walls and thus small $R$, leading to potentially large non-Gaussiatities. Let us choose the largest mass one may consistently consider in a low energy effective field theory, $m\sim 1$, in order to derive an upper bound on non-Gaussianities. 
Since $V(R)\sim C_0$, we get
\begin{eqnarray}
R\sim\sqrt{2C_0}\,.
\end{eqnarray}
For such a hole the non-linearity parameter becomes independent of $C_0$, leading to the bound $f_{NL}\gg -81$ for $\mu\ll \mu_{max}$, which is still in excess of the observational limits\footnote{$\mu= \mu_{max}$ leads to $f_{NL}= -81$; note that $\mu\ll \mu_{max}$ corresponds to $\Delta P_\zeta \ll P_\zeta^{\mbox{\tiny COBE}}$, which we assumed in our computations, leading to $f_{NL}\gg -81$. If we want to saturate $\Delta P_\zeta = P_\zeta^{\mbox{\tiny COBE}}$, a large negative contribution to $f_{NL}$ in excess of the PLANCK limit results; a correction to the given expression of $f_{NL}$ of order one arises in this case, which could be computed as in \cite{Battefeld:2011yj}, which would however not change this conclusions.  } if one pushes $\mu$ close to $\mu_{max}$  ; due to the large, negative contribution to the Bi-spectrum, it is impossible to make the additional contribution to the power-spectrum $\Delta P_\zeta$ in (\ref{DeltaPzeta}) the dominant one if $m\sim 1$, justifying in retrospective our working hypothesis in this paper that slow roll dynamics in $\phi_1$ is responsible for the amplitude of the power-spectrum and the scalar spectral index. If the hole is shallower, for instance because $m\ll 1$ in the quadratic potential considered here, one could make the additional contribution to the power-spectrum the dominant one without violating the upper bond non-Gaussianities due to PLANCK, similar to the mechanism considered in \cite{Battefeld:2011yj}. In this case the current fine tuning needed to get a red spectral tilt may be alleviated. We leave this interesting possibility to future studies.

To achieve $f_{NL}\geq -1$ for $m\sim 1$, the impact parameter has to satisfy $\mu/R\leq 0.984$, which is only a modest reduction of the allowed interval in (\ref{boundonmu}) by about $1$\%. Thus, almost all encounters with holes that are as shallow as the ones considered here are consistent with current bounds on non-Gaussianities. Indeed, for natural values of $\mu/R\sim 0.5$, we get an unobservable small
\begin{eqnarray}
f_{NL}\sim -0.000097\,,
\end{eqnarray}
consistent with the absence of non-Gaussianities in current measurements. 

It should be noted that $g^{\prime\prime}$ always leads to a negative $f_{NL}$ since the trajectory enters a hole on the inflationary plateau\footnote{$f_{NL}$ would be positive if the trajectory were to fall off a ridge.}. If an observation of positive, primordial local non-Gaussianity were to be made, this simple type of inflation on the landscape would need to be complemented by additional mechanisms, such as a strongly curved inflationary trajectory or a non-trivial fast foll regime after inflation but before the hole is encountered (falling off a ridge etc.). However, in the random potentials considered in \cite{Frazer:2011br,McAllister:2012am} the general prediction for inflation near a saddle point is a  ``reasonably'' straight trajectory, such that $f_{NL}$ remains unobservable small in most cases, see e.g. \cite{Frazer:2011br,McAllister:2012am}.
Therefore, the absence of a positive, local $f_{NL}$ is a prediction in this class of models that could however be avoided if additional non-trivial dynamics is present.

\subsection{A Smooth Landscape}

While anthropic arguments can mitigate the fine-tuning needed to drive inflation on the string theoretical landscape (i.e.~finding a flat enough saddle point as well as a final resting place with a small but positive vacuum energy), they are of little use when it comes to non-Gaussanities. The prediction of negative, potentially large, local non-Gaussianities, which are not observed, puts constraints onto the type of landscapes that we should consider for inflationary model building. Concretely, the inflationary field space should be relatively smooth in the sense that features, such as the hole where inflation ends, are shallow: their width needs to be considerably wider than their height in natural units, 
\begin{eqnarray}
\Delta \varphi_{\text{feature}} > \Delta V^{1/n}\quad  \text{with} \quad n\geq 2\,. 
\end{eqnarray}
If features are sharper, the end of inflation needs to occur at a rare, shallow hole, putting us at a special place in the landscape. Based on the Copernican principle, we should abandon such scenarios and deduce that the potential on the landscape is smooth in the above sense. Given this new insight into properties of the field space, we can justify in retrospective the modelling of random landscapes by means of truncated Fourier series with smaller amplitudes the higher the frequency, as in \cite{Tegmark:2004qd,Frazer:2011tg,Frazer:2011br,Battefeld:2012qx}.

We compute the tri-spectrum in appendix \ref{app:trispectrum}, which follows in a similar manner, but does not lead to additional insights at the time of writing, because observational bounds are strongest for the bi-spectrum.

\subsection{Recap via an Analogy}

Let us recap our findings by means of an analogy to a soccer game (representing slow roll inflation); we want to figure out where in the world a game has been played, a region with steep valleys and mountains, like Switzerland, or a reasonably flat region, like the Netherlands (what properties does the string-theoretical landscape posses?). To play a game somewhere in the world, one needs to find a flat region (the inflationary plateau) in the first place; once such a place is found (antropics), the game lasts for a while (SR inflation); at the end of the game, the ball is kicked out of the field, and rolls down in the landscape surrounding the field, until it finds its final resting place - this place needs to be above the water-level, as the ball is lost otherwise (the VEV of the inflation today needs to have a small but positive CC; anthropics). Besides this property, the local minimum that the ball lands into, should not be special (Copernican principle). We  pick up the ball today  and look where it ended up: a wide or narrow hole (measurements of e.g.~non-Gaussianities): we find that the ball has landed in a wide hole, $\text{width} \sim \sqrt{\text{depth}}$ as opposed to $\text{width} \sim \text{depth}$ (a steep hole leads to observationally ruled out non-Gaussianities). Due to this property, we deduce that the game has most likely been played in a region like the Netherlands as opposed to Switzerland (we conclude that the landscape is smooth in the above sense).

\section{Conclusion}
Based on the proposition of inflation on a higher dimensional landscape of string theory, we considered small field inflation near a saddle point, terminated by the trajectory leading into a hole on the inflationary plateau. The encounters with a sufficiently flat plateau as well as a final resting place with a positive, but small cosmological constant are rare, requiring the use of anthropic arguments in conjunction with eternal inflation. Guided by prior numerical analyses of inflation on random potentials, we focused on a straight trajectory and expanded the potential near the saddle; parameters are determined by comparison of the predicted power-spectrum (caused by the inflaton) with observations, which leave one parameter undetermined, i.e.~the inflationary energy scale $V_{\text{inf}}$; the latter is, however, bounded from above by the requirement of small-field inflation. 

To this known, simple and analytically tractable inflationary setup, we added a hole of radius $R$ and depth $V_{\text{inf}}$ on the plateau, to terminate inflation and stabilize all light fields subsequently. Since the end-of-inflation hypersurface is curved, additional contributions to correlation functions arise as perturbations in the perpendicular fields cause slight changes in the volume expansion rate, leading in turn to curvature fluctuations. Based on the $\delta N$-formalism, we computed the dominant additional contribution to the power-spectrum, brought forth by the perpendicular fields. Requiring this contribution to be sub-dominant compared to the inflationary one (which we justified subsequently, based on our computation of non-Gaussianities), we derived an upper bound on the impact parameter,  $0<\mu\ll \mu_{\text{max}}<R$.

We followed with a computation of non-Gaussianities, paying particular attention to the non-linearity parameter $f_{NL}$, measuring  the amplitude of the Bi-spectrum in the local limit. Due to the convex end-of-inflation hypersurface, we find $f_{NL}<0$, which is severely constrained by current observations.

To make further progress, we needed to specify the radius of the hole. Based on the Copernican principle, we argued that the features of the hole, such as the width to depth ratio, should be representative of features on the landscape\footnote{The smooth inflationary plateau is a rare occurrence, which requires the use of anthropic arguments; however, the features of the hole are not subject to any anthropic constraints beyond the requirement that all fields should be heavy and that the value of the potential at the minimum should be small but positive.}. Assuming that the width of the hole is comparable to the height in natural units leads to an unacceptable large, negative $f_{NL}$ for the vast majority of allowed impact parameters. Hence, such landscapes are ruled out.

The next simplest hole has a simple parabolic potential inside, entailing $R\propto \sqrt{V_{\text{inf}}}$. We find that the non-linearity parameter is in excess of the allowed limits in about $1\%$ of the allowed interval of the impact parameter, while being unobservable small for natural values of $\mu\sim R/2$. Thus, such smooth landscapes, as well as even smoother ones, are consistent with current observations.

Additional non-Gaussianities could be sourced by other mechanisms, such as turns of the inflationary trajectory, but it is known that the simplest models do not predict measurable non-Gaussianities. Thus, if a positive, primordial  $f_{NL}$ is detected, the simplest models of inflation on such smooth multi-field landscapes are ruled out. The absence of non-Gaussianities in the current PLANCK data release is therefore consistent with these models.

One important application of our results is the retrospective justification of approximating the landscape by a random potential based on a truncated Fourier series, whose coefficients are exceedingly suppressed, the higher the frequency of the Fourier mode. Such potentials, reminiscent of a beginners ski slope without sharp cliffs, have been used extensively in the literature, see e.g.~\cite{Tegmark:2004qd,Frazer:2011tg,Frazer:2011br,Battefeld:2012qx}.

\acknowledgments
We would like to thank the APC (Paris) for hospitality during the final stages of writing and J.~Frazer for discussions motivating this study. 

\appendix
\section{The Tri-spectrum \label{app:trispectrum}}
The non-linearity parameters of the tri-spectrum are bounded by observations of the CMBR to \cite{Ade:2013tta}
\begin{eqnarray}
\tau_{NL}&<& 2800\,,\label{limittaunl}
\end{eqnarray}
at the $2\sigma$-level and \cite{Desjacques:2009jb,Vielva:2009jz,Smidt:2010sv,Smidt:2010ra}
\begin{eqnarray}
|g_{NL}|&\lesssim& 10^6\label{limitgnl}\,;
\end{eqnarray}
they can be computed in the $\delta N$-formalism to 
\begin{eqnarray}
\tau_{NL}=\frac{N_{ij}N^{ik}N^jN_k}{\left(N_lN^l\right)^3}\quad ,\quad
g_{NL}= \frac{25}{54}\frac{N_{ijk}N^iN^jN^k}{\left(N_lN^l\right)^3} \,.
\end{eqnarray}
Note that dynamics during preheating can alter significantly the value of these parameters \cite{Leung:2013rza}, but as a general rule of thumb, one may use the values derived in the $\delta N$-formalism to estimate the order of magnitude.
In our case, they become
\begin{eqnarray}
\tau_{NL}&=&\frac{\frac{C_0^4}{C_1^4}\frac{g^{\prime 2}}{\left[1+g^2\frac{3C_3}{C1})\right]^4}\left[\frac{-6C_3}{C_1}\frac{1}{\left(1+g^2\frac{3C_3}{C1}\right)}gg^{\prime 2}+g^{\prime\prime}\right]^2}{\left(P_\zeta\frac{4\pi^2}{H^2}\right)^3}\,, \label{tauNL}\\
g_{NL}&=&\frac{25}{54}\frac{\frac{C_0^4 g'^3}{C_1^4\left(1+g^2\frac{3C_3}{C1}\right)^4}\left(\left[g^{\prime 3}\left(\frac{12C_3g^2}{C_1\left(1+g^2\frac{3C_3}
{C1}\right)}-1\right)-3gg^\prime g^{\prime\prime}\right]\frac{6C_3}{C_1\left(1+g^2\frac{3C_3}{C_1}\right)} +
g^{\prime\prime\prime}\right)}
{\left(P_\zeta\frac{4\pi^2}{H^2}\right)^3}\,.\label{gNL}
\end{eqnarray}
Using the same line of reasoning that lead to (\ref{fNLsimplified}) and
\begin{eqnarray}
g^{\prime\prime\prime}|_{\phi_2^*=0}=\frac{3}{R^2}\frac{\frac{\mu}{R}}{\left(1-\frac{\mu^2}{R^2}\right)^{5/2}}\,,
\end{eqnarray}
we can simplify (\ref{tauNL}) and (\ref{gNL}) to
\begin{eqnarray}
\tau_{NL}&=&A_3f_3\left(x\right) - A_4f_4\left(x\right) \sqrt{\frac{C_0}{C_0^{\text{max}}}}\frac{1}{R}+ A_5f_5\left(x\right)\frac{C_0}{C_0^{\text{max}}}\frac{1}{R^2}\,,\\
g_{NL}&=&A_6f_6(x) - A_7f_7\left(x\right) \sqrt{\frac{C_0}{C_0^{\text{max}}}}\frac{1}{R}+ A_8f_8\left(x\right)\frac{C_0}{C_0^{\text{max}}}\frac{1}{R^2},
\end{eqnarray}
where $A_i$ are again constants, $x=\mu/R$, and $f_i(x)$ are simple functions given by combining derivatives of $g$. 
To be concrete, they are given by
\begin{eqnarray}
A_3&\equiv& \frac{6^2}{(2\pi)^6P^3_{\zeta}3^3}\frac{C_0^{\text{max}\,7}}{C_1^{\text{max}\,6}}C_3^{\text{max}\,2}\Delta\phi^2_{\text{max}}\left(1+3\Delta\phi^2_{\text{max}}\frac{C_3^{\text{max}}}{C_1^{\text{max}}}\right)^{-6} \,,\\
A_4&\equiv&\frac{12}{(2\pi)^6P^3_{\zeta}3^3}
\frac{C_0^{\text{max}\,7}}{C_1^{\text{max}\,5}}C_3^{\text{max}}
\Delta\phi_{\text{max}}\left(1+3\Delta\phi^2_{\text{max}}\frac{C_3^{\text{max}}}{C_1^{\text{max}}}\right)^{-5} \,,\\
A_5&\equiv & \frac{1}{(2\pi)^6P^3_{\zeta}3^3}
\frac{C_0^{\text{max}\,7}}{C_1^{\text{max}\,4}}\left(1+3\Delta\phi^2_{\text{max}}\frac{C_3^{\text{max}}}{C_1^{\text{max}}}\right)^{-4} \,,\\
\nonumber A_6 &\equiv&\frac{25}{54} \frac{6}{(2\pi)^6P^3_{\zeta}3^3}\frac{C_0^{\text{max}\,7}}{C_1^{\text{max}\,5}}C_3^{\text{max}}\left(12\left(1+3\Delta\phi^2_{\text{max}}\frac{C_3^{\text{max}}}{C_1^{\text{max}}}\right)^{-1}\Delta\phi_{\text{max}}^2\frac{C_3^{\text{max}}}{C_1^{\text{max}}}-1\right)\\
&&\times \left(1+3\Delta\phi^2_{\text{max}}\frac{C_3^{\text{max}}}{C_1^{\text{max}}}\right)^{-5}\,,\\
A_7 &\equiv&\frac{25}{54} \frac{18}{(2\pi)^6P^3_{\zeta}3^3}\frac{C_0^{\text{max}\,7}}{C_1^{\text{max}\,5}}C_3^{\text{max}}\Delta\phi_{\text{max}}\left(1+3\Delta\phi^2_{\text{max}}\frac{C_3^{\text{max}}}{C_1^{\text{max}}}\right)^{-5} \,,\\
A_8 & \equiv& \frac{25}{54}\frac{1}{(2\pi)^6P^3_{\zeta}3^2}\frac{C_0^{\text{max}\,7}}{C_1^{\text{max}\,4}}\left(1+3\Delta\phi^2_{\text{max}}\frac{C_3^{\text{max}}}{C_1^{\text{max}}}\right)^{-4} \,,
\end{eqnarray}
and
\begin{eqnarray}
f_3(x)&\equiv& \frac{x^6}{(1-x^2)^3} \,,\\
f_4(x)&\equiv& \frac{x^4}{(1-x^2)^{7/2}} \,,\\
f_5(x)&\equiv& \frac{x^2}{(1-x^2)^4} \,,\\
f_6(x)&\equiv& \frac{x^6}{(1-x^2)^3} \,,\\
f_7(x)&\equiv& \frac{x^4}{(1-x^2)^{7/2}} \,,\\
f_8(x)&\equiv& \frac{x^4}{(1-x^2)^4} \,.
\end{eqnarray}

Let us focus on the case of a smooth landscape with $R\sim \sqrt{2C_0}$, that lead to a bi-spectrum consistent with current bounds. Evidently both $\tau_{NL}$ and $g_{NL}$ become model independent, i.e.~independent of $C_0$, like $f_{NL}$. Taking $\mu/R=0.984$ as the impact parameter (tuning at the percent level), corresponding to $f_{NL}=-1$, we get $\tau_{NL}\approx 1540$ and $g_{NL}\approx 2070$, which are both below the current limits in (\ref{limittaunl}) and (\ref{limitgnl}). Interestingly, the bounds on $\tau_{NL}$ and $f_{NL}$ from PLANCK \cite{Ade:2013tta} lead to similar constraints on the impact parameter. More natural values of the impact parameter $\mu/R\sim 1/2$, corresponding to $f_{NL}\sim -0.000097$, lead to
\begin{eqnarray}
\tau_{NL}\sim 0.0013\,,\\
g_{NL}\sim 0.00046 \,,
\end{eqnarray}
which are unobservable small. 
Note that $\tau_{NL}\gg (f_{NL}6/5)^2$, since the dominant contribution of the bi- and tri-spectrum originate from $\phi_2$, whereas the power-spectrum is given by fluctuations in $\phi_1$ (this is similar to the limit $\Xi\ll 1$ in \cite{Battefeld:2011yj}). Thus, higher order correlation functions are in principle easier to detect than in models that saturate the Suyama-Yamagutchi bound \cite{Suyama:2007bg}.

\end{document}